\documentclass[a4paper,11pt]{article}
\pdfoutput=1 \usepackage{jheppub_2} 
\usepackage[T1]{fontenc} 
\usepackage[utf8]{inputenc}
\usepackage{color,subfigure}
\usepackage{graphicx,multirow,soul,url,amsmath,amsfonts,amssymb,mathrsfs,amsfonts}
\usepackage[all]{hypcap}
\usepackage{url}
\usepackage{bbm}
\usepackage{cancel}
\usepackage{subfigure}
\usepackage{slashed}
\usepackage[dvipsnames]{xcolor}
\usepackage{multicol, blindtext}
\usepackage[section]{placeins}
\usepackage[version=3]{mhchem}
\usepackage{caption}
\usepackage{lipsum}
\usepackage{hyperref}
\usepackage{float}
\usepackage{mattens}
\usepackage{mathtools}
\usepackage{footnote}
\usepackage{amsmath,amssymb,amsthm,amsxtra,overpic,bbm,bm,epsfig}
\usepackage{color,subfigure}
\usepackage[splitrule]{footmisc}
\usepackage{bbold}
\usepackage{amsmath}
\usepackage[normalem]{ulem}

\newcommand{\beq}{\begin{equation}}
\newcommand{\eeq}{\end{equation}}

\newcommand{\SU}{\,{\rm SU}}

\newcommand{\U}{\,{\rm U}}

\newcommand{\matrixel}[3]{\left< #1 \vphantom{#2#3} \right|
 #2 \left| #3 \vphantom{#1#2} \right>} % for Dirac matrix elements
\let\baraccent=\= % rename builtin command \= to \baraccent

\notoc 

\title{\boldmath The QCD Axion and Unification}
\author[a]{Pavel Fileviez P\'erez,}
\author[b]{Clara Murgui}
\author[a]{and Alexis D. Plascencia}
\affiliation[a]{Physics Department and Center for Education and Research in Cosmology and Astrophysics (CERCA), 
Case Western Reserve University, Cleveland, OH 44106, USA}
\affiliation[b]{Departamento de F\'isica Te\'orica, IFIC, Universitat de Valencia-CSIC, 
E-46071, Valencia, Spain}

\emailAdd{pxf112@case.edu}
\emailAdd{clara.murgui@ific.uv.es}
\emailAdd{alexis.plascencia@case.edu}

\abstract{The QCD axion is one of the most appealing candidates for the dark matter in the Universe. In this article, we discuss the possibility to predict the axion mass in the context of a simple renormalizable grand unified theory where the Peccei-Quinn scale is determined by the unification scale. In this framework, the axion mass is predicted to be in the range $m_a \simeq (3 - 13) \times 10^{-9} \  \rm{eV}$. We study the axion phenomenology and find that the ABRACADABRA and CASPEr-Electric experiments will be able to fully probe this mass window.}

\begin{document} 

\maketitle
\flushbottom

\newpage 

\section{Introduction}

The need for cold dark matter to explain the features of our Universe is well established today.  
The QCD axion~\cite{Peccei:1977hh,Wilczek:1977pj,Weinberg:1977ma} is one of the most popular candidates for the cold dark matter in the Universe~\cite{Preskill:1982cy,Abbott:1982af,Dine:1982ah}, which was introduced to solve the so-called strong CP problem.
For reviews about axions see Refs.~\cite{Raffelt:1990yz,Dine:2000cj,Sikivie:2006ni,Kim:2008hd,Jaeckel:2010ni,Graham:2015ouw,Irastorza:2018dyq}. Using the Peccei-Quinn (PQ) mechanism~\cite{Peccei:1977hh} and after QCD confinement the axion mass can be written as~\cite{Weinberg:1977ma}
\begin{equation*}
m_a^2 \simeq \frac{m_u m_d}{(m_u + m_d)^2} \frac{m_\pi^2 f_\pi^2}{f_{a}^2},
\end{equation*} 
where $f_{\pi}$ is the pion constant, $m_\pi$ is the pion mass, $m_u$ is the mass of the up quark, $m_d$ the mass of the down quark and $f_{a}$ is the Peccei-Quinn breaking scale. This relation implies that the axion mass can only be predicted if the scale $f_a$ can be computed. See Ref.~\cite{diCortona:2015ldu,Gorghetto:2018ocs} for an improved relation between the axion mass and $f_a$.

There are two simple mechanisms where the interaction between the axion and the gluons can be understood. Firstly, there is the KSVZ mechanism~\cite{Kim:1979if,Shifman:1979if} in which the
axion is the imaginary part of the scalar field that couples to new vector-like quarks, and once these heavy quarks are integrated out, the coupling between the axion and the gluons $a(x) G \tilde{G}$ is generated. Secondly, there is the DFSZ mechanism~\cite{Zhitnitsky:1980tq, Dine:1981rt} 
where there are two Higgs doublets which couple to the Standard Model (SM) fermions and a complex scalar singlet. In this scenario the axion is mainly the imaginary part of the singlet, 
it couples to the SM fermions through scalar mixing, and the $a(x) G \tilde{G}$ coupling is generated after integrating out the SM quarks.
However, these simple models do not predict the PQ scale, and hence, the axion mass and its interactions with the SM gauge bosons and fermions cannot be determined.

In this work, we study the idea of predicting the PQ scale in the context of grand unified theories (GUTs) where the unification of the Standard Model gauge interactions can be understood. 
In this context, if the same field breaking the GUT symmetry contains the axion, the PQ and unification scales will be related. This idea was first pointed out by M. B. Wise, H. Georgi and S. Glashow in Ref.~\cite{Wise:1981ry}. In this article, we investigate this idea and the predictions in the context of a simple renormalizable grand unified theory where the PQ and seesaw scales are determined by the unification scale. We focus on the scenario where the PQ symmetry is broken before inflation.

We discuss the implementation of the Peccei-Quinn mechanism in the context of grand unified theories and show that in a simple renormalizable $\SU(5)$ theory the KSVZ mechanism can be implemented.
We find that in the context of Adjoint $\SU(5)$~\cite{Perez:2007rm}, the PQ and seesaw scales are determined by the unification scale; also, a realistic framework for all fermion masses including neutrino masses can be achieved.
In this theory, the PQ scale is given by
$$f_a = \frac{1}{10}\sqrt{\frac{6}{5\pi \alpha_{\rm GUT} }} M_{\rm GUT},$$
where $M_{\rm GUT}$ is the unification scale and $\alpha_{\rm GUT}$ is the value of the gauge couplings at the unification scale.
Once we take into account experimental constraints on the proton lifetime and LHC constraints, the axion mass is predicted in the range 
$$m_a \simeq (3 - 13) \times  10^{-9} \  \rm{eV}.$$ 
We discuss in detail the testability of this theory at current and future axion experiments, and investigate the implications for collider and proton decay experiments.
We find that the ABRACADABRA~\cite{Kahn:2016aff}  and CASPEr-Electric~\cite{Budker:2013hfa} axion experiments will be able to fully probe the predicted mass window.

This article is organized as follows, we discuss the implementation of the PQ mechanism in grand unified theories in section~\ref{sec:PQUnification}. In section~\ref{sec:Axion}, we show how to implement the 
PQ mechanism in the context of Adjoint $\SU(5)$, we discuss the unification of gauge interactions, the constraints from proton decay experiments, the predictions for the axion mass and the predictions for different axion experiments. 
Our main results are summarized in section~\ref{sec:Summary}.

%%%%%%%%%%%%%%%%%%%%%%%%%
\section{Predicting the PQ Scale and Unification}
\label{sec:PQUnification}
%%%%%%%%%%%%%%%%%%%%%%%%

To investigate the relation between the PQ and unification scales we will focus on minimal unified theories which are based on the $\SU(5)$ gauge symmetry.
In the minimal $\SU(5)$ theory proposed by H. Georgi and S. Glashow~\cite{Georgi:1974sy} the SM matter is unified in ${\bf{\bar{5}}}$ and ${\bf{10}}$, the Higgs sector is composed 
of ${\bf{5_H}}$ and ${\bf{24_H}}$, and all the gauge fields live in ${\bf{24_V}}$.
It is well-known that this theory is not realistic because: a) the values of the gauge 
couplings at the low scale cannot be reproduced, b) it predicts the relation $Y_e=Y_d^T$, where $Y_e$ and $Y_d$ are the Yukawa matrices for charged leptons and down quarks, respectively. 
Last but not least, c) neutrinos are massless as in the SM. In order to have a consistent relation between the fermion masses, $Y_e \neq Y_d^T$, a 
Higgs in the ${\bf{45_H}}$ representation can be introduced~\cite{Georgi:1979df}. In this scenario, both the ${\bf{5_H}}$ and the ${\bf{45_H}}$ must couple to matter in the same manner. This Higgs sector can provide the 
relevant degrees of freedom which allow us to have gauge coupling unification in agreement with the experimental values of the gauge couplings. For a recent study see 
Ref.~\cite{Perez:2016qbo}.

In order to predict the PQ scale from unification, the same field that breaks the grand unifying group must simultaneously break the PQ symmetry. In $\SU(5)$ the symmetry is broken once 
the ${\bf{24_H}}$ acquires a vacuum expectation value (VEV); when this field is charged under $\U(1)_{\rm PQ}$ it becomes a complex field and the axion will be contained in the imaginary part of the singlet inside the  ${\bf{24_H}}$.
This simple idea is very appealing because it allows us to write down 
a simple GUT model with a dark matter candidate as well. In the previous discussion, we have mentioned that in renormalizable $\SU(5)$ the ${\bf{5_H}}$ and ${\bf{45_H}}$ must 
couple to matter in the same way, then if we implement the PQ mechanism these Higgses must have the same PQ charge. This implies that the DFSZ mechanism cannot be implemented in this minimal renormalizable theory.

For the implementation of the KSVZ mechanism we require extra matter fields coupled to the ${\bf{24_H}}$ in order to generate the $a(x) G \tilde{G}$ coupling. We can consider the fermionic ${\bf{24}}$~\cite{Ma:1998dn,Bajc:2006ia}, which also generates the neutrino masses through the type I~\cite{Minkowski:1977sc,Mohapatra:1979ia,GellMann:1980vs,Yanagida:1979as} and type III~\cite{Foot:1988aq} seesaw mechanisms. Consequently, the KSVZ mechanism can be implemented in the context of Adjoint $\SU(5)$~\cite{Perez:2007rm} and the axion mass can be predicted in a realistic renormalizable grand unified theory. 
See e.g. Refs.~\cite{Co:2016vsi,Boucenna:2017fna,DiLuzio:2018gqe,Ernst:2018bib} for the implementation of the PQ mechanism in the context of GUTs. 

\begin{center}
\begin{table}[t]
\centering
\begin{tabular}{cccc}
\hline
$Y_e \neq Y_d^T$  & $M_\nu \neq 0$                                                                 & {\rm{PQ}}                                                                 & {\rm{Unification}}                                                                    \\ \hline
${\bf{45_H}}$    & \begin{tabular}[c]{@{}l@{}} ${\bf{1_i}} $ \\ \\ ${\bf{10_H}}$  \\ \\ ${\bf{15_H}}$  \\ \\ ${\bf{24}}$  \end{tabular} & \begin{tabular}[c]{@{}l@{}} $\textcolor{red}{\times}$ \\  \\  $\textcolor{red}{\times}$\\ \\  $\textcolor{red}{\times}$\\ \\  $\textcolor{blue}{\checkmark}$ \end{tabular} & \begin{tabular}[c]{@{}l@{}} $\textcolor{blue}{\checkmark}$\\ \\      $\textcolor{blue}{\checkmark}$\\ \\      $\textcolor{blue}{\checkmark}$\\ \\     $\textcolor{blue}{\checkmark}$\end{tabular} \\ \hline \\
${\bf{5^{'}}}$ + ${\bf{\bar{5}^{'}}}$ & \begin{tabular}[c]{@{}l@{}} ${\bf{1_i}}$ \\ \\ ${\bf{15_H}}$ \\ \\ ${\bf{24}}$ \end{tabular}          & \begin{tabular}[c]{@{}l@{}} $\textcolor{blue}{\checkmark}$\\ \\  $\textcolor{blue}{\checkmark}$\\ \\  $\textcolor{blue}{\checkmark}$\end{tabular}         & \begin{tabular}[c]{@{}l@{}} $\textcolor{red}{\times}$\\ \\      $\textcolor{red}{\times}$\\ \\      $\textcolor{red}{\times}$\end{tabular}     \\
\hline       
\end{tabular}
\caption{\label{theories} Minimal renormalizable gauge theories based on $\SU(5)$. The third column indicates whether the PQ symmetry leading to the axion can be implemented. The fourth column indicates whether gauge unification can be achieved.}
\end{table}
\end{center}

Recently, the authors in Ref.~\cite{DiLuzio:2018gqe} studied the axion mass predictions in a non-renormalizable $\SU(5)$ theory. The main differences between the study in Ref.~\cite{DiLuzio:2018gqe} and our study are the following: a) Our model is a simple realistic renormalizable model based on SU(5) where the PQ mechanism can be implemented without requiring Planck scale suppressed operators, while in Ref.~\cite{DiLuzio:2018gqe} these operators are needed in order to achieve realistic masses for the charged fermions, split the mass of the fermions in the ${\bf{24}}$ and give mass to more than one neutrino. Generically, this type of operators could break the PQ symmetry. b) In our model, we have implemented the KSVZ mechanism to generate the axion mass, while in Ref.~\cite{DiLuzio:2018gqe} they have two mechanisms, the KSVZ and the DFSZ.

In Table~\ref{theories} we present the different renormalizable theories based on $\SU(5)$ with the minimal number of representations where the Peccei-Quinn mechanism could be implemented. As we have discussed above, if we use the ${\bf{45_H}}$ to correct the charged fermion masses, only the theory with the fermionic ${\bf{24}}$ can be realistic. In the case where the ${\bf{5^{'}}}$ + ${\bf{\bar{5}^{'}}}$ vector-like fermions are used to correct the charged fermion masses and considering the different possibilities to generate neutrino masses, we need to add more representations in order to achieve unification. See for example Ref.~\cite{FileviezPerez:2018dyf} for the study of $\SU(5)$ theories with extra ${\bf{5^{'}}}$ + ${\bf{\bar{5}^{'}}}$ vector-like matter. Notice that when the PQ mechanism is implemented, the splitting in the ${\bf{24}}$ is small and unification cannot be achieved in agreement with bounds from proton decay, see Ref.~\cite{FileviezPerez:2018dyf} for details.
Moreover, new Higgses in the fundamental representation could be added, as it was done in Ref.~\cite{Wise:1981ry}, to implement the DFSZ mechanism but this also requires extra representations. 
Then, it can be argued that the theory we will investigate in this article is the realistic renormalizable theory based on $\SU(5)$ with the minimal number of representations where the Peccei-Quinn mechanism can be realized.

%
%%%%%%%%%%%%%%%%%%%%%%
\section{The QCD Axion and Adjoint SU(5)}
\label{sec:Axion}
%%%%%%%%%%%%%%%%%%%%%%
In Adjoint $\SU(5)$~\cite{Perez:2007rm} the KSVZ mechanism can be implemented if the different representations are charged under the PQ symmetry as follows,
\begin{eqnarray}
&& \bar{5} \to e^{- 3 i  \theta} \bar{5}, \hspace{1.6cm}  10 \to  e^{ + i \theta} 10, \hspace{1.2cm}   5_H \to e^{ - 2 i \theta} 5_H, \nonumber \\
&& 24_H \to e^{ - 10 i \theta} 24_H,  \hspace{0.5cm}   45_H \to e^{ - 2 i \theta} 45_H, \hspace{0.5cm}  24 \to e^{ + 5 i \theta} 24. \nonumber
\end{eqnarray}
We note that the relative PQ charges among all representations are fixed by the consistency of the theory.
Then, we can achieve a realistic relation between the charged fermion masses using the following Yukawa interactions
\begin{equation}
{\cal{L}} \supset 10 \ 10 \left( Y_1 \ 5_H \ + \ Y_2 \ 45_H \right) + 10 \ \bar{5} \left( Y_3 \ 5_H^* \ + \ Y_4 \ 45_H^* \right) + \rm{h.c.},
\end{equation}
from which the charged fermion mass matrices read as
\begin{eqnarray}
M_U &=& \frac{1}{\sqrt{2}} \left(  2 \left( Y_1 + Y_1^T \right) v_5 - 4 \left( Y_2 - Y_2^T \right) v_{45} \right), \\
M_E &=&  \frac{1}{2} \left( Y_3 v_5^* - 6 Y_4 v_{45}^* \right), \\
M_D &=& \frac{1}{2} \left( Y_3^T v_5^* + 2 Y_4^T v_{45}^* \right),
\end{eqnarray}
where $ \left< 5_H \right> = v_5 /\sqrt{2}$, $\left< 45_H \right>^{15}_1=\left< 45_H \right>^{25}_2=\left< 45_H \right>^{35}_3=v_{45}/\sqrt{2}$ and $\left< 45_H \right>^{45}_4=-3v_{45}/\sqrt{2}$.

In this theory, the neutrino masses are generated through the type I~\cite{Minkowski:1977sc,Mohapatra:1979ia,GellMann:1980vs,Yanagida:1979as}, type III~\cite{Foot:1988aq} and 
the colored seesaw~\cite{FileviezPerez:2009ud} mechanisms using the interactions
\begin{equation}
{\cal{L}} \supset h_1 \ \bar{5} \ 24 \ 5_H \ + \ h_2  \ \bar{5} \ 24 \ 45_H  + \lambda {\rm{Tr}} \left[ 24^2 \ 24_H \right] + \rm{h.c.}.
\end{equation}
The neutrino mass matrix is given by
\begin{equation}
M_\nu = M_\nu^{\rm I} + M_\nu^{\rm III} +  M_\nu^{\rm cs},
\end{equation}
where $M_\nu^{\rm I}$, $M_\nu^{\rm III}$, and $M_\nu^{\rm cs}$ are the type I, type III and colored seesaw contributions, respectively.
See Ref.~\cite{Perez:2007rm,Emmanuel-Costa:2013gia} for the explicit form of the neutrino mass matrix and further details.

Now, in order to implement the KSVZ mechanism we use the fact that in this theory the axion lives in the 
${\bf{24_H}}$, i.e. $${\bf{24_H}} \supset \frac{1}{\sqrt{2}} |\Sigma_0| \ e^{\frac{i a(x)}{v_{\Sigma}}},$$
where $v_{\Sigma}$ is the VEV of $\Sigma_0$. This simple idea is appealing because the Peccei-Quinn scale can be predicted from gauge coupling unification since the same field breaking the GUT symmetry contains the axion field.
See Appendix~\ref{sec:AppRep} for the explicit form of all fields in the theory and Refs.~\cite{Perez:2008ry,Blanchet:2008cj,Blanchet:2008ga} for the 
study of the phenomenological aspects of Adjoint SU(5).
\begin{figure}[tbp]
\centering
\includegraphics[width=0.495\linewidth]{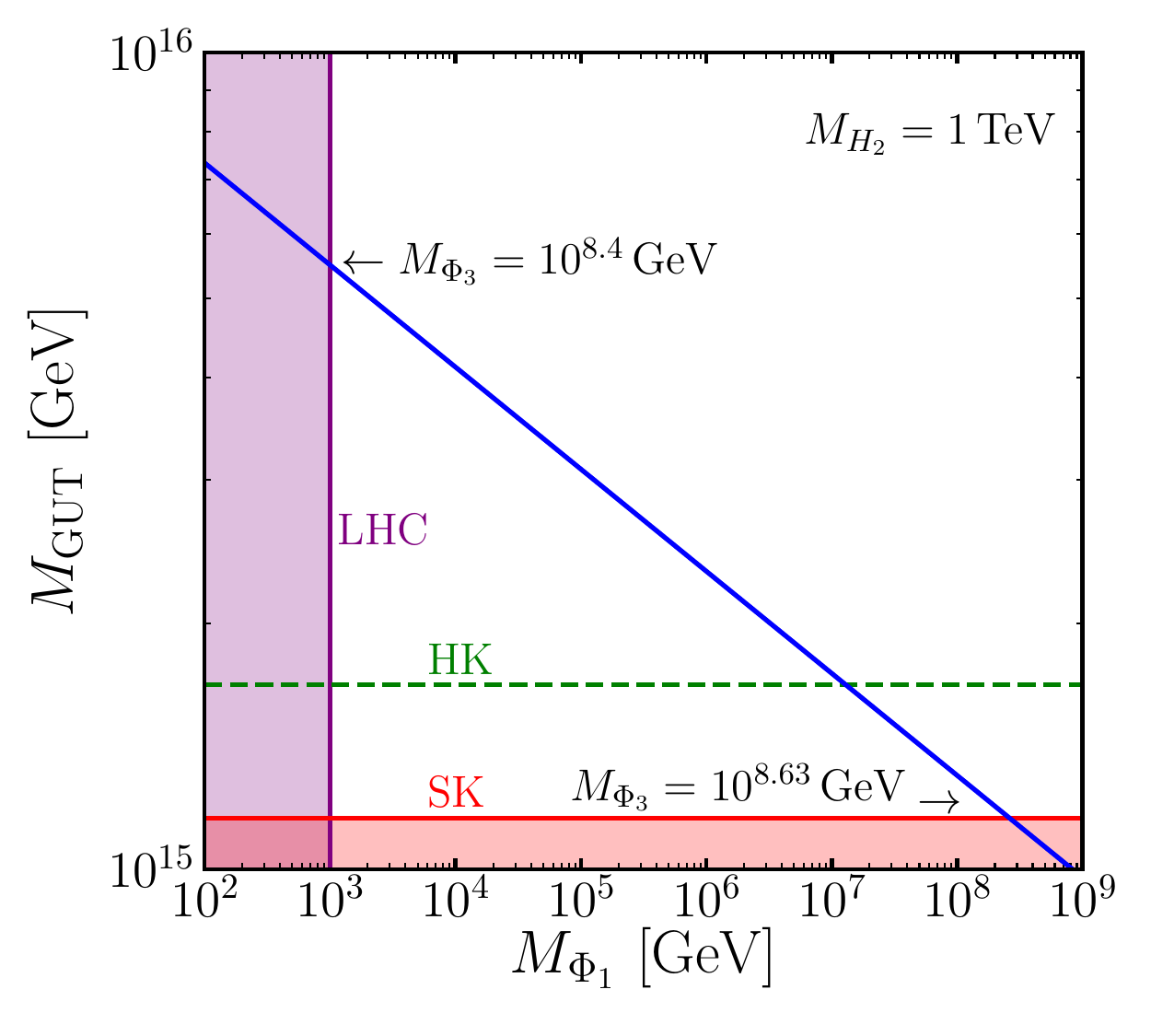}
\includegraphics[width=0.495\linewidth]{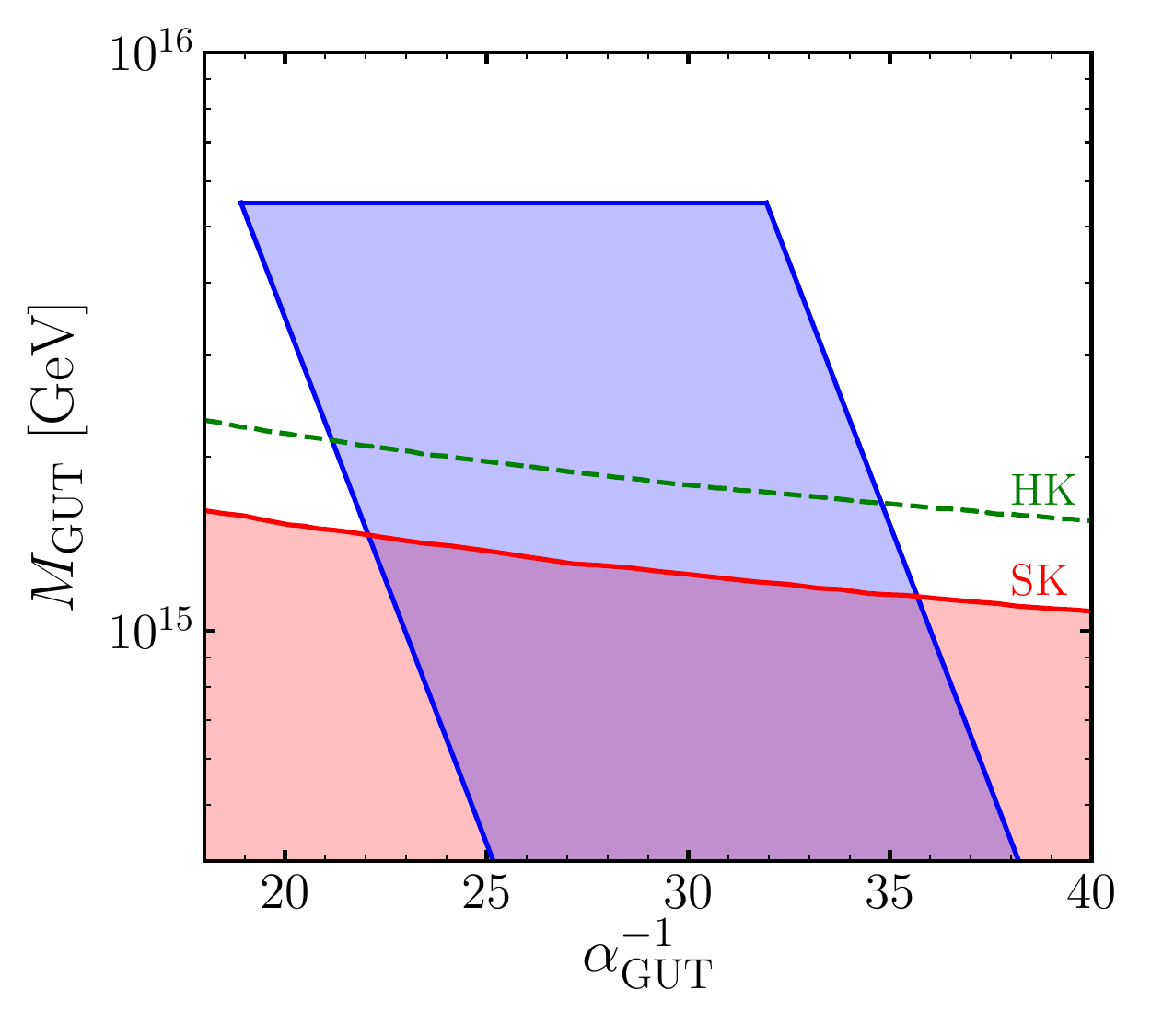}
\caption{\small{ {\textit{Left Panel}}: Predictions for the GUT scale, $M_{\rm GUT}$, as a function of $M_{\Phi_1}$. We fix the mass of the second Higgs doublet $H_2$ living in the ${\bf 45_H}$ to $M_{H_2} = 1$ TeV. 
The purple shaded region shows the current LHC bound on the scalar colored octet $\Phi_1 \sim (8,2,1/2)$, $M_{\Phi_1}>1$ TeV~\cite{Hayreter:2017wra, Miralles:2019uzg}. The region shaded in red shows the current experimental limit on the proton lifetime from the Super-Kamiokande collaboration, $\tau(p \to K^+ \bar{\nu}) > 5.9 \times 10^{33}$ years~\cite{Abe:2014mwa}, and the green dashed line shows the projected bound for the Hyper-Kamiokande collaboration, $\tau(p\to K^+ \bar{\nu}) > 2.5 \times 10^{34}$ years~\cite{Yokoyama:2017mnt}. When applying the bounds from proton lifetime we take the most conservative limit which corresponds to taking $M_{\rho_3}=10^{15}$ GeV. The mass of the $\Phi_3\sim (3,3,-1/3)$ varies between the range $10^{8.4}-10^{8.63}$ GeV from left to right. {\textit{Right Panel:}}  Relation between the GUT scale and $\alpha_{\rm GUT}$. The blue shaded region leads to gauge unification, the width of this band is determined by varying from $M_{\rho_3} = 10^{4.51}$ GeV (left limit determined by the BBN bounds, as discussed in Appendix~\ref{sec:AppBBN}) to $M_{\rho_3} = 10^{15} $ GeV (right limit determined by the perturbative bound on the seesaw scale). Here we use $M_{H_2} = 1$ TeV as in the left panel.}}
\label{fig:GUT}
\end{figure}
%
%%%%%%%%%%%%%%%%%%%%%%%%%%%%%
\subsection{Unification Constraints and Proton Decay}
\label{sec:Unification}
%%%%%%%%%%%%%%%%%%%%%%%%%%%%

\begin{figure}[h]
\centering
\includegraphics[height=8cm]{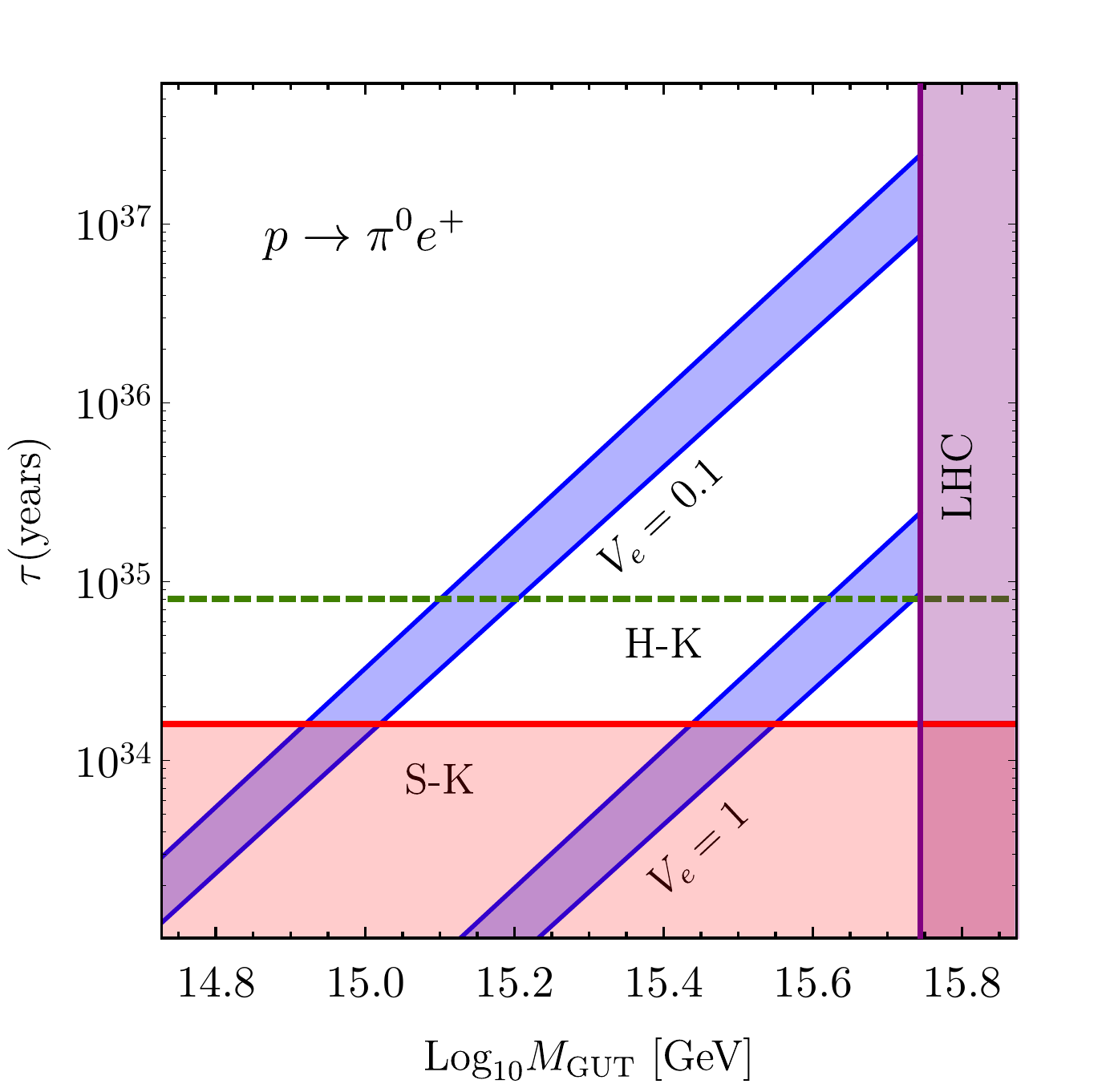}
\caption{Predictions for the proton decay lifetime for the channel $p \to \pi^0 e^+$. The different blue bands show different choices of the magnitude of the unknown mixing matrix $V_e$. The width of the blue bands is determined by the allowed values of $\alpha_{\rm GUT}$ as we vary $M_{\rho_3} \subset [ 10^{4.51}, 10^{15}]$ GeV. The red shaded area shows the excluded parameter space by the bound on the proton decay lifetime, $\tau ( p\to \pi^0 e^+) > 1.6 \times 10^{34}$ years~\cite{Miura:2016krn}, from the Super-Kamiokande collaboration. The green dashed line shows the projected bound on proton decay lifetime from the Hyper-Kamiokande collaboration, i.e. $\tau(p \to \pi^0 e^+) > 8 \times 10^{34}$ years~\cite{Yokoyama:2017mnt}.}
\label{ptopi0eplus}
\end{figure}
\begin{figure}[h]
\centering
\includegraphics[height=7cm]{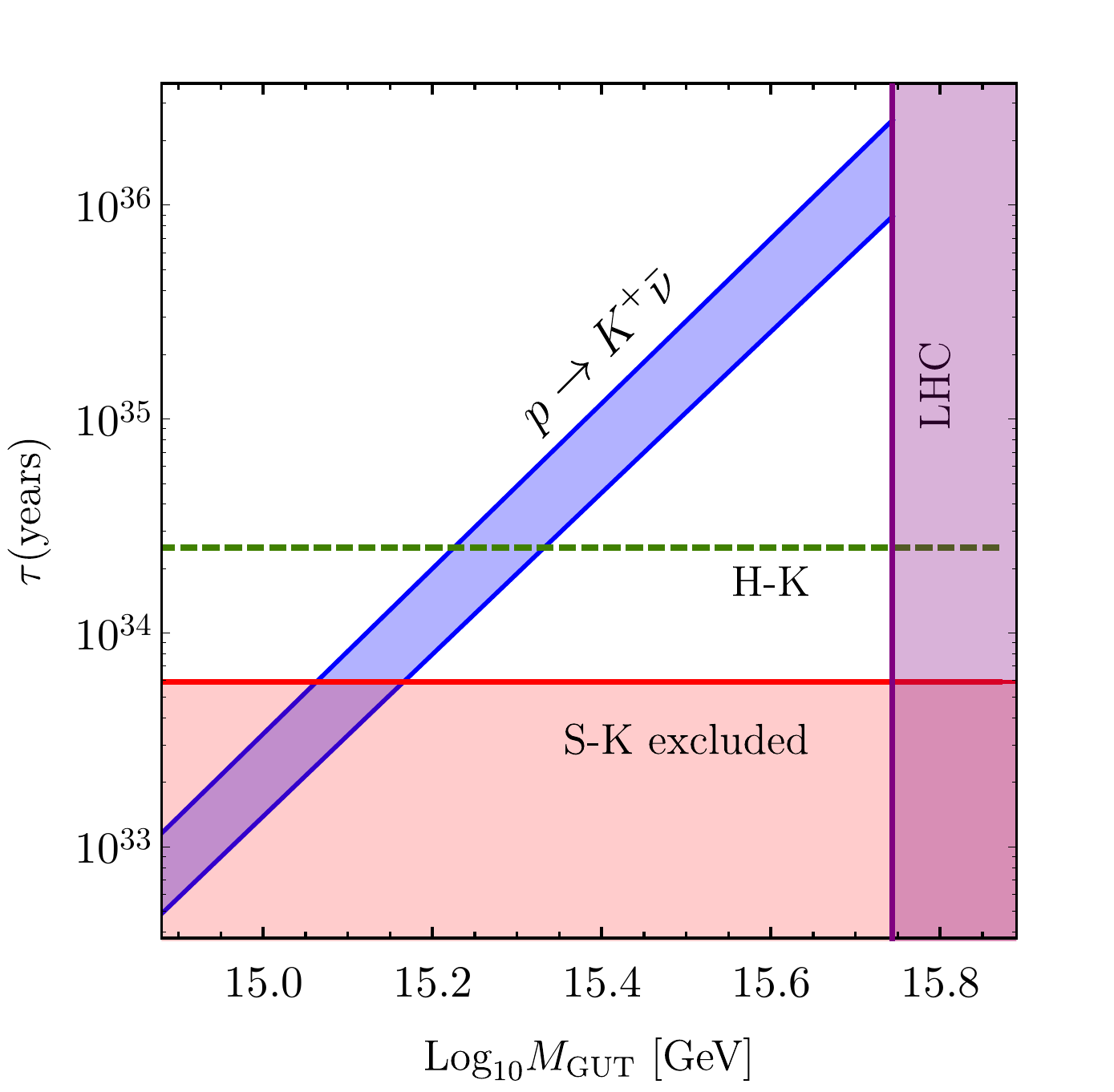}
\includegraphics[height=7cm]{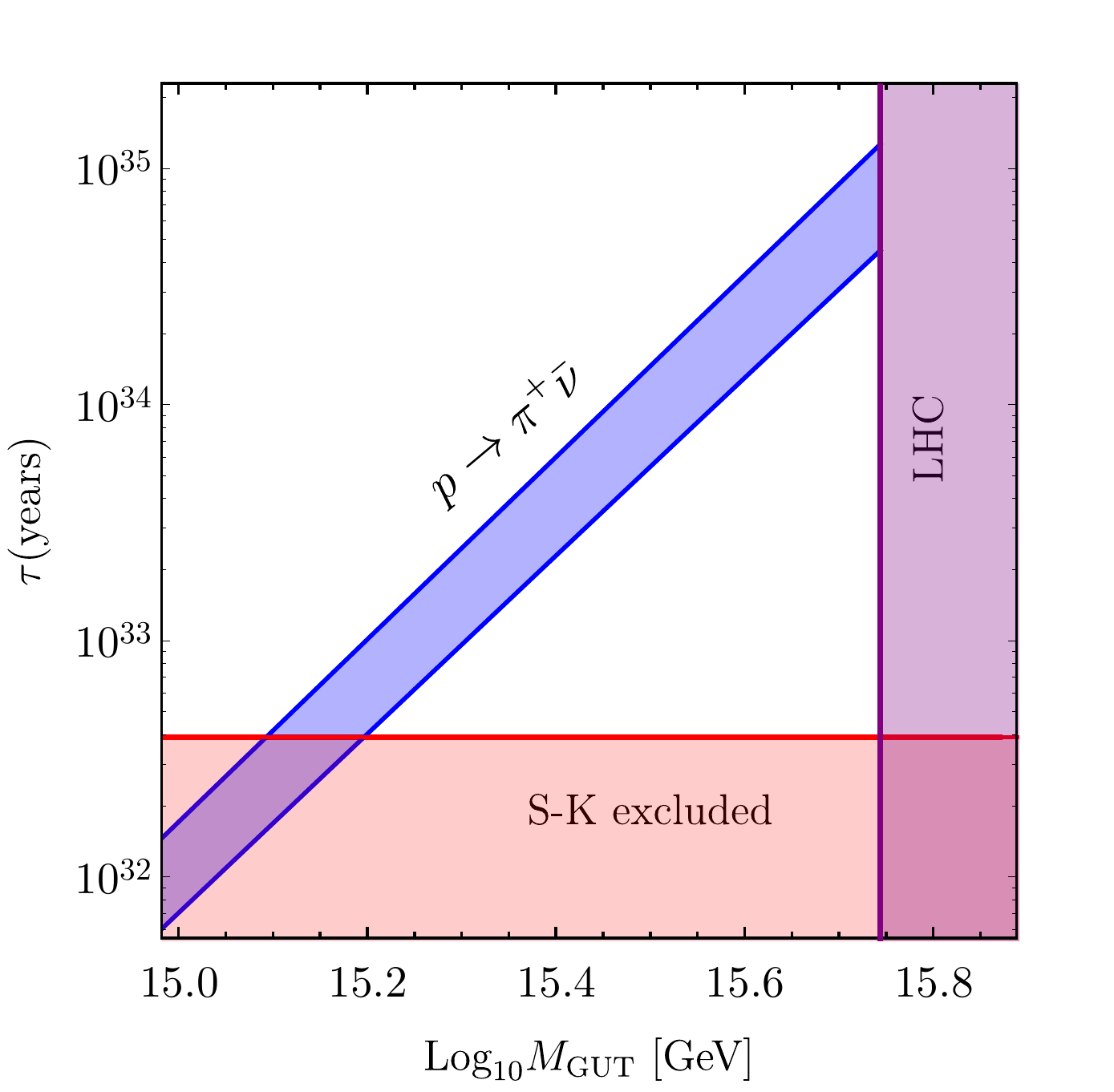}
\caption{Predictions for the proton decay channels $p \to K^+ \bar{\nu}$ and $p \to \pi^+ \bar{\nu}$. The blue bands show the possible range of variation of $\alpha_{\rm GUT}$, according to $M_{\rho_3} \subset [10^{4.51}, 10^{15}]$ GeV. The red shaded areas show the excluded region by the Super-Kamiokande constraints, $\tau(p \to K^+ \bar{\nu}) > 5.9 \times 10^{33}$ years \cite{Abe:2014mwa} and $\tau(p \to \pi^+ \bar{\nu})> 3.9 \times 10^{32}$ years \cite{Abe:2013lua}. The green dashed line shows the projected bound by the Hyper-Kamiokande collaboration on the $p \to K^+ \bar{\nu}$ channel, $\tau(p\to K^+ \bar{\nu}) > 2.5 \times 10^{34}$ years~\cite{Yokoyama:2017mnt}.}
\label{fig:nudecays}
\end{figure}

In order to predict the PQ scale we need to understand the predictions for the GUT scale.
The renormalization group equations (RGEs) for the gauge couplings can be written as
\begin{equation}
\frac{1}{\alpha_i (M_Z)} - \frac{1}{\alpha_i (\mu)} = \frac{1}{2 \pi} b_i^{\rm SM} \ {\rm{ln}} \frac{\mu}{M_Z} + \frac{1}{2 \pi} \sum_I b_{iI} \ \Theta (\mu - M_I) \, {\rm{ln}} \frac{\mu}{M_I},
\end{equation}
where $i=1,2,3$, $b_1^{\rm SM}=41/10$, $b_2^{\rm SM}=-19/6$ and $b_3^{\rm SM}=-7$. In Appendix~\ref{sec:AppBeta} we list the different contributions to the beta functions in Adjoint $\SU(5)$.
Using the above relations, the following equations must be satisfied in order to achieve unification at the one-loop level~\cite{Giveon:1991zm},
\begin{eqnarray}
\frac{B_{23}}{B_{12}} &=& \frac{5}{8} \left(  \frac{\sin^2 \theta_W(M_Z) - \alpha(M_Z) / \alpha_s(M_Z)}{3/8 - \sin^2 \theta_W (M_Z)}\right)= 0.717, \label{unificationCONS1} \\
{\rm{ln}}  \frac{M_{\rm GUT}}{M_Z} &=& \frac{16 \pi}{5 \alpha (M_Z)} \left(   \frac{3/8 - \sin^2 \theta_W (M_Z)}{B_{12}} \right)=\frac{184.95}{B_{12}},
\label{unificationCONS2}
\end{eqnarray}
where the parameters $B_{ij}$ are defined in Appendix~\ref{sec:AppBeta}. Here we have used the experimental values $\sin^2 \theta_W (M_Z)=0.23122$, $\alpha (M_Z)=1/127.955$, and $\alpha_s (M_Z)=0.1181$ at the $M_Z$ scale~\cite{Tanabashi:2018oca}.
It is well-known that in the minimal $\SU(5)$ theory unification cannot be achieved in agreement with the experimental values of the gauge couplings at the low scale 
since $B_{23}/B_{12} < 0.6$. 

In the context of Adjoint $\SU(5)$, in the ${\bf 45_H}$ representation there are only three fields that help achieve unification: for $\Phi_3$ and $H_2$, the lighter they are the larger the fraction $B_{23}/B_{12}$ can be. The field $\Phi_1$ does not help to increase the later ratio; however, it helps towards unification indirectly since it gives a negative contribution to $B_{12}$, thus increasing the GUT scale. This is necessary, as we will show, to avoid the strong proton decay experimental constraints. With the help of these three fields, unification can be achieved in agreement with experimental results. The rest of fields in the ${\bf 45_H}$ will be assumed to be, without loss of generality, at the GUT scale since we are interested in finding the allowed parameter space by proton decay bounds.

For the case of the $\bf{24}$ representation, the small mass splitting between the masses of their fields is fixed and given by 
\begin{eqnarray}
M_{\rho_0}=\frac{1}{3} M_{\rho_3}, \quad 
M_{\rho_8}=\frac{2}{3} M_{\rho_3}, \quad
M_{\rho_{(3,2)}}=M_{\rho_{(\bar{3},2)}}=\frac{1}{6} M_{\rho_3},
\end{eqnarray}
which is a direct consequence of ${\bf 24}$ being charged under the PQ-symmetry.  We note that the unification constraints do not depend on the global mass scale of the fields in the ${\bf 24}$, i.e. $M_{\rho_3}$; they are only sensitive to their mass splitting. 

In the left panel in Fig.~\ref{fig:GUT} we show the predictions of the GUT scale for the different values of $M_{\Phi_1}$. We fix the mass of the second Higgs doublet $H_2$ living in the ${\bf 45_H}$ to $M_{H_2} = 1$ TeV.
The purple shaded region shows the current LHC bound on the scalar colored octet $\Phi_1 \sim (8,2,1/2)$, $M_{\Phi_1 }> 1$ TeV~\cite{Hayreter:2017wra, Miralles:2019uzg}. The region shaded in red shows the current experimental limit on the proton lifetime from the Super-Kamiokande collaboration, $\tau(p \to K^+ \bar{\nu}) > 5.9 \times 10^{33}$ years~\cite{Abe:2014mwa}, and the green dashed line shows the projected bound for the Hyper-Kamiokande collaboration, $\tau(p\to K^+ \bar{\nu}) > 2.5 \times 10^{34}$ years~\cite{Yokoyama:2017mnt}.
When applying the bounds from proton lifetime we take the most conservative limit which corresponds to taking $M_{\rho_3}=10^{15}$ GeV.
The mass of the $\Phi_3\sim (3,3,-1/3)$ varies between the range $10^{8.4}-10^{8.63}$ GeV from left to right. 
This result is quite general since $H_2$ has a small contribution to the running of the gauge couplings.
For the choice $M_{H_2} = 1$ TeV, the allowed range for the GUT scale is $M_\text{GUT} = \left[ 10^{15.06}- 10^{15.74} \right]$ GeV, the lower limit is determined by the bound on the proton lifetime from Super-Kamiokande on the $p \to K^+ \bar{\nu}$ channel and the upper one by the collider bound on the mass of $\Phi_1$.

The allowed values for $\alpha_\text{GUT}$ depend on the values for $M_{\rho_3}$, which define a possible range of values for the GUT gauge coupling.
The fields in the $\bf{24}$ must be at the seesaw scale or below in order to generate realistic values for the neutrino masses. In the right panel in Fig.~\ref{fig:GUT}, we present the relation between the 
GUT scale and $\alpha_{\rm GUT}$. The blue shaded region leads to gauge unification, the width of this band is determined by varying from $M_{\rho_3} = 10^{4.51}$ GeV (left limit) 
to $M_{\rho_3} = 10^{15} $ GeV (right limit), and we have used $M_{H_2} = 1$ TeV. The lower limit $M_{\rho_3} = 10^{4.51}$ GeV is determined by the bound from Big Bang Nucleosynthesis (BBN) discussed in Appendix~\ref{sec:AppBBN}, 
while $M_{\rho_3} = 10^{15} $ GeV is the perturbative bound using the seesaw relation for neutrino masses.

In the left panel in Fig.~\ref{fig:GUT} we have demonstrated that, in order to have unification in agreement with all experimental constraints, the mass of the $\Phi_3$ field must lie in the range $10^{8.4}-10^{8.63}$ GeV. 
This is possible only if the coupling of $\Phi_3$ to matter, $Y_2$, is suppressed because this field mediates proton decay. It is well-known 
that if $Y_2 \sim Y_U$ we need $M_{\Phi_3} > 10^{12}$ GeV to suppress proton decay~\cite{Nath:2006ut}. 
Fortunately, in this theory we can have that $Y_2 \ll Y_1 \approx Y_U$, and hence, we can have a consistent scenario for fermion masses where the mass matrix 
for up-quarks $M_U$ is approximately symmetric. In the case when $M_U$ is symmetric, we can make clean predictions for the proton 
decay channels with anti-neutrinos~\cite{FileviezPerez:2004hn}. See Appendix~\ref{sec:AppProton} for more details. 

In Fig.~\ref{ptopi0eplus}  we show the predictions for the proton decay lifetime for the channel $p \to \pi^0 e^+$. The decay width for this channel is given by 
\begin{eqnarray}
\Gamma (p \to \pi^0 e^+)&=& \frac{\pi m_p}{2} \frac{\alpha_{\rm GUT}^2}{M_{\rm GUT}^4} A^2  \left(  \left| (V_2^{11} +V_{UD}^{11}(V_2 V_{UD})^{11})\right |^2 + \left |V_3^{11} \right |^2 \right) \left |\matrixel{\pi^0}{(ud)_R u_L}{p}\right|^2  \nonumber \\
&=&   \frac{\pi m_p}{2} \frac{\alpha_{\rm GUT}^2}{M_{\rm GUT}^4}  A^2  V_e^2  \left |\matrixel{\pi^0}{(ud)_R u_L}{p}\right|^2.  
\end{eqnarray}
Here we parametrize the magnitude of the unknown mixings with $V_e$, and we show the predictions in Fig.~\ref{ptopi0eplus} for two different values: $V_e = 0.1$ and $V_e=1$. See the Appendix~\ref{sec:AppProton} for more details. The blue bands in Fig.~\ref{ptopi0eplus} include the possible different values of $\alpha_{\text{GUT}}$ according to the choice of $M_{\rho_3} \subset [ 10^{4.51}, 10^{15}]$ GeV. The red shaded area shows the excluded parameter space by the bound on the proton decay lifetime, $\tau ( p\to \pi^0 e^+) > 1.6 \times 10^{34}$ years~\cite{Miura:2016krn}, from the Super-Kamiokande collaboration. The green dashed line shows the projected bound on proton decay lifetime from the Hyper-Kamiokande collaboration, i.e. $\tau(p \to \pi^0 e^+) > 8 \times 10^{34}$ years~\cite{Yokoyama:2017mnt}. Since the lifetime for this channel is a function of an unknown matrix $V_e$, 
we proceed to investigate the channel with antineutrinos that can give us a more conservative bound on the GUT scale.

In Fig.~\ref{fig:nudecays} we show the predictions for the proton decay channels $p \to K^+ \bar{\nu}$ and $p \to \pi^+ \bar{\nu}$. 
We have discussed above that we need to work in the scenario where $Y_2 \ll Y_1$. In this case $M_U \simeq M_U^T$ and the different proton decay channels can be written as
\begin{eqnarray}
 \Gamma (p \to K^+ \bar{\nu}) &\simeq&  \frac{\pi m_p}{2} \frac{\alpha_{\rm GUT}^2}{M_{\rm GUT}^4}  \left( 1- \frac{m_{K^+}^2}{m_p^2}\right)^2 A^2  \nonumber \\ &\times & \left( \left |V_{\rm CKM}^{11}  
 \matrixel{K^+}{(us)_R d_L}{p} \right |^2 + \left | V_{\rm CKM}^{12}  
 \matrixel{K^+}{(ud)_R s_L}{p} \right|^2 \right), \\ 
 \Gamma (p \to \pi^+ \bar{\nu}) &\simeq&  \frac{\pi m_p}{2} \frac{\alpha_{\rm GUT}^2}{M_{\rm GUT}^4}  A^2 \left|V_{\rm CKM}^{11} \matrixel{\pi^+}{(du)_R d_L}{p}\right |^2,
\end{eqnarray}
where $V_{\rm CKM}^{ij}$ are the elements of the $V_{\rm CKM}$ matrix. See the Appendix~\ref{sec:AppProton} for more details. Since in this case the decay lifetime is a function of known parameters we can make a better prediction for these channels.
The blue bands in Fig.~\ref{fig:nudecays} show the possible range of variation of $\alpha_{\rm GUT}$, according to $M_{\rho_3} \subset [10^{4.51}, 10^{15}]$ GeV. The red shaded areas show the excluded region by the Super-Kamiokande constraints, $\tau(p \to K^+ \bar{\nu}) > 5.9 \times 10^{33}$ years \cite{Abe:2014mwa} and $\tau(p \to \pi^+ \bar{\nu})> 3.9 \times 10^{32}$ years \cite{Abe:2013lua}. The green dashed line shows the projected bound by the Hyper-Kamiokande collaboration on the $p \to K^+ \bar{\nu}$ channel, $\tau(p\to K^+ \bar{\nu}) > 2.5 \times 10^{34}$ years~\cite{Yokoyama:2017mnt}. Therefore, in the rest of this article we will use the range of $M_{\rm GUT}$ allowed by the experimental bounds on proton decay into antineutrinos as shown in Fig.~\ref{fig:nudecays}, which are the most conservative allowed values for $M_{\rm GUT}$.

%%%%%%%%%%%%%%%%%%%%%
\subsection{Axion Mass and Couplings}
%%%%%%%%%%%%%%%%%%%%
\begin{figure}[tbp]
\centering
\includegraphics[width=.9\textwidth]{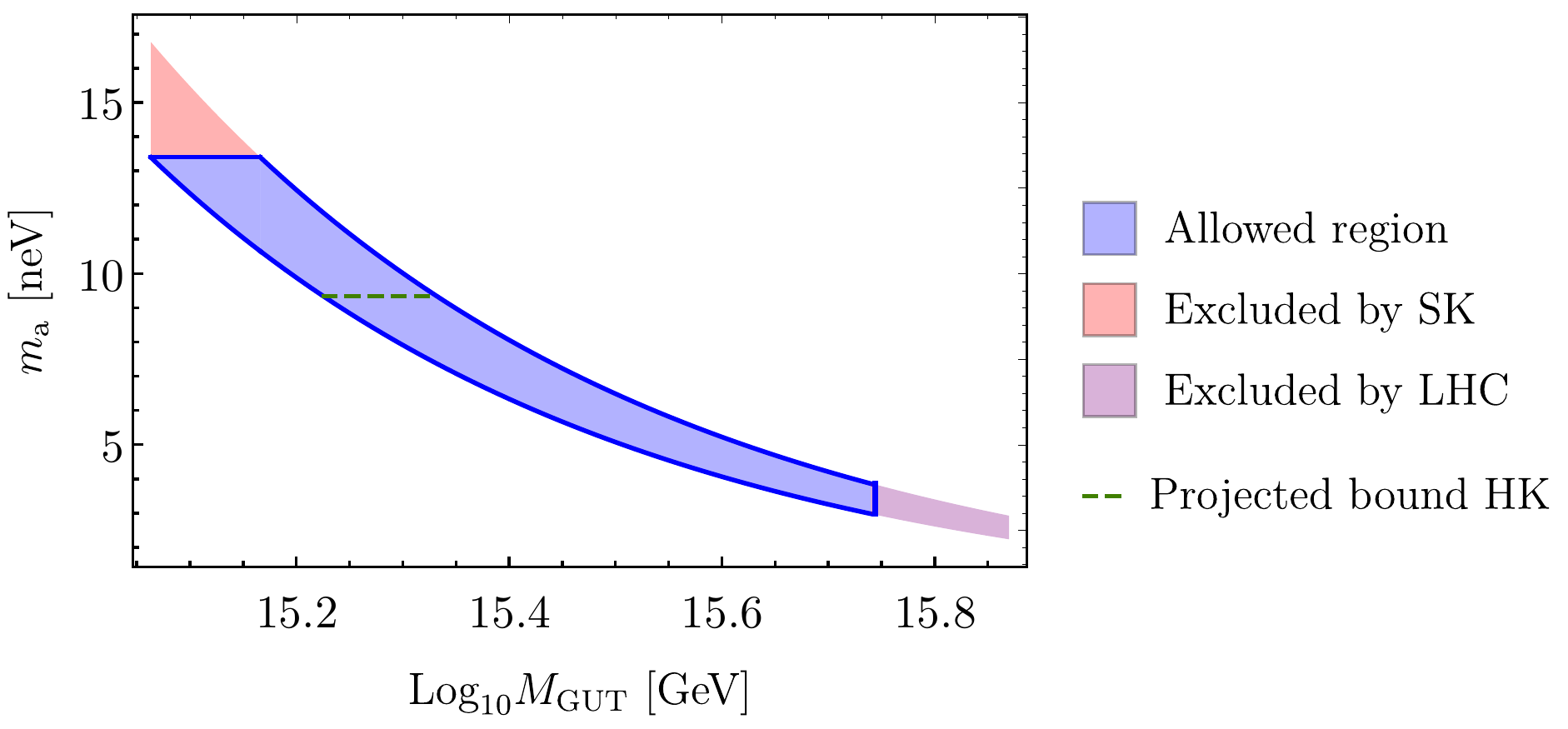}
\caption{\small{Axion mass as a function of the GUT scale. The blue band shows the parameter space allowed by unification constraints and proton decay. The red (purple) shaded area shows the excluded parameter space by the proton decay bound (LHC bound on $M_{\Phi_1}$). The dashed green line gives the projected sensitivity for Hyper-Kamiokande.}}
\label{axionmass}
\end{figure}
In this theory, the axion is only coupled to the heavy fermions in the ${\bf{24}}$, and hence, we can use the KSVZ mechanism. The interaction comes from the Yukawa term between ${\bf{24}}$ and ${\bf{24_H}}$
\begin{equation}
\label{eq:Yukawa}
{\cal L} \supset \lambda \ \text{Tr}\left[ 24^2 24_H \right] + {\rm h.c.}.
\end{equation}
Once the singlet inside the ${\bf{24_H}}$  acquires a non-zero vacuum expectation value, we can replace ${\bf{24_H}}$ in the above equation by
\beq
\label{eq:24vev}
 24_H \to  \frac{v_\Sigma}{\sqrt{15}} \  {\rm{diag}} (-1, -1, -1, 3/2, 3/2) \, e^{ia(x)/v_\Sigma},
\eeq
where $a(x)$ corresponds to the axion field, and we find that
\beq
\mathcal{L} \supset \frac{ \lambda}{\sqrt{15}} \, v_\Sigma \, e^{ia(x)/v_\Sigma} \left( - {\rm Tr}[ \rho_8  \rho_8 ] +  \frac{1}{2} {\rm Tr}[\rho_{(\bar{3},2)} \rho_{(3,2)}] + \frac{3}{2} {\rm Tr}[\rho_3 \rho_3] + 
\frac{1}{2} \rho_0^2 \right) + {\rm h.c.}.
\eeq

Following the method proposed by Fujikawa~\cite{Fujikawa:1980eg}, the coupling to the gluons can be written as
\beq
\mathcal{L} \supset \frac{\alpha_3} {8 \pi} \frac{a}{v_\Sigma} N G_{\mu\nu} \tilde{G}^{\mu\nu} =  \frac{\alpha_3} {8 \pi} \frac{a}{f_a} G_{\mu\nu} \tilde{G}^{\mu\nu},
\eeq
where $N$ depends on the chiral fields that have been rotated in order to absorb the axion field, and in our case $N=5$.
The coupling of the axion to the gluons is generated by the rotations of the colored fields in the ${\bf 24}$ representation.
Hence, the relation between $f_a$ and the GUT scale is given by
\beq
f_a \equiv \frac{v_\Sigma}{N} = \frac{v_\Sigma}{5} = \sqrt{\frac{6}{5\pi \alpha_{\rm GUT}}} \frac{M_{\rm GUT}}{10},
\eeq
where we have used $M_V\!=\!M_{\rm GUT}\!=\! \sqrt{5/6} \, g_{\rm GUT} \, v_\Sigma$. 
This is one of the most important results of this paper which allows us to predict the axion mass once we know the allowed values for $M_{\rm GUT}$ and $\alpha_{\rm GUT}$.
In order to show our numerical results we will use the following relation between $m_a$ and $f_a$~\cite{diCortona:2015ldu}
\begin{equation}
m_a=5.70(7) \, \times 10^{-6} \ \text{eV} \left(\frac{10^{12}\text{ GeV}}{f_a}\right). 
\end{equation} 
As demonstrated in Sec.~\ref{sec:Unification}, in Adjoint $\SU(5)$ a small range for the GUT scale and $\alpha_{\rm GUT}$ can be predicted, this leads to the following prediction for the axion mass 
\beq
 m_a =  \left( 2.98-13.4 \right) \times 10^{-9} \ {\rm eV}. 
\eeq

In Fig.~\ref{axionmass} we show the predictions for the axion mass as a function of the GUT scale. The blue band shows the parameter space allowed by unification constraints and proton decay. The red (purple) shaded area shows the excluded parameter space by the proton decay bound (LHC bound on $M_{\Phi_1}$). This prediction is crucial to understand the testability of this theory at axion experiments because it tells us the specific range for the axion mass where a signal could be expected. In the allowed mass window, an initial misalignment angle of $\theta_i \approx 10^{-2}$ \cite{Preskill:1982cy,Abbott:1982af,Dine:1982ah} must be assumed in order for the axion field to saturate the dark matter relic abundance $\Omega_{\rm DM} h^2 = 0.1200 \pm 0.0012$ \cite{Aghanim:2018eyx}.
\begin{figure}[tbp]
\centering
\includegraphics[width=.85\textwidth]{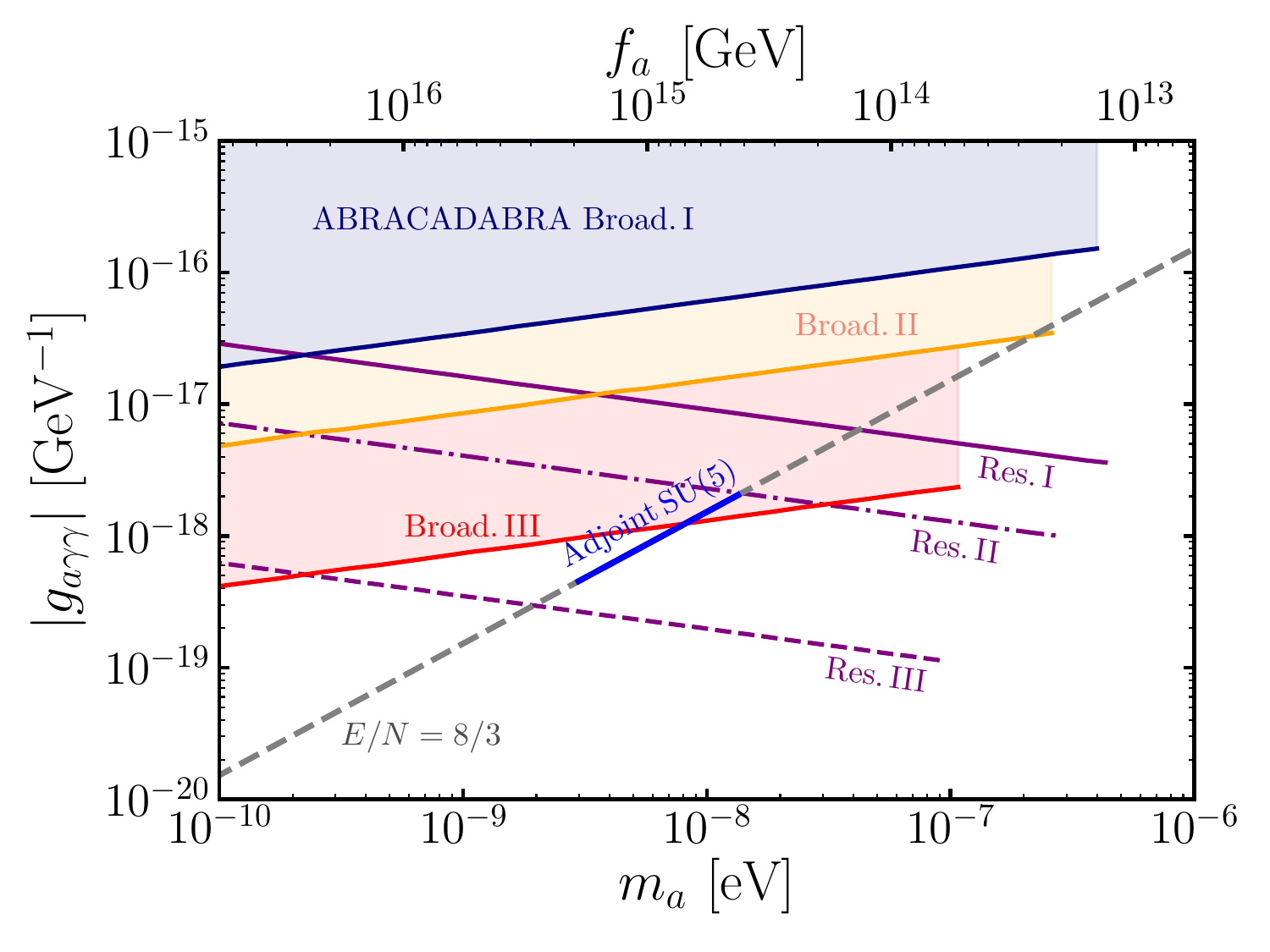} 
\caption{\small{The axion-photon coupling versus the axion mass. The blue line corresponds to the axion mass prediction in Adjoint $\SU(5)$. The regions shaded in blue, orange and red correspond to the projected sensitivities for Phase I, II and III of the ABRACADABRA experiment \cite{Kahn:2016aff} using the broadband approach. The purple lines correspond to the projected sensitivities for the three different phases in the resonant approach. }}
\label{gagammagamma}
\end{figure}

We proceed to study the phenomenology of the axion in the predicted mass window. The coupling between the axion and photons can be written as
\beq
\mathcal{L} \supset -\frac{g_{a\gamma\gamma}}{4} a F_{\mu\nu} \tilde{F}^{\mu\nu},
\eeq
with the effective coupling $g_{a\gamma\gamma}$ given by
\beq
\label{eq:gagamma}
g_{a\gamma\gamma} = \frac{\alpha_{\tiny{\rm EM}}} {2\pi f_a} \left(\frac{E}{N}  - 1.92(4) \right)  = \frac{\alpha_{\tiny{\rm EM}}} {2\pi f_a} \left(\frac{8}{3}  - 1.92(4) \right),
\eeq
where second term is the contribution from non-perturbative effects from the axion coupling to QCD and has been computed at NLO in Ref.~\cite{diCortona:2015ldu}. The first term in Eq.~\eqref{eq:gagamma} is given by the anomaly coefficient for the electromagnetic fields and $E/N\!=\!8/3$ since we are working with fermions in a complete representation of $\SU(5)$. 

The ABRACADABRA experiment~\cite{Kahn:2016aff} is expected to be sensitive to the axion to photon coupling for very light axions, $m_a \lesssim 10^{-6}$ eV. In Fig.~\ref{gagammagamma} we present the axion to photon coupling together with the projected reach for ABRACADABRA, as given in Ref.~\cite{Kahn:2016aff}. The regions shaded in different colors correspond to the projected sensitivity for the broadband operating mode, while the purple lines show the ones for the resonant operating mode. The blue solid line corresponds to the prediction in Adjoint $\SU(5)$. It is of great interest that the third stage of the broadband mode will be able to probe a portion of this mass range, while the third stage of the resonant operating mode could fully probe the predicted mass window.

At low energies, the axion field has also a coupling to the electric dipole moment (EDM) operator for nucleons
\beq
\mathcal{L} \supset - \frac{i}{2} g_{aD} \, a ( \bar{\psi}_N  \, \sigma^{\mu\nu} \gamma^5 \psi_N ) F_{\mu \nu},
\eeq
where $\sigma^{\mu\nu}\!= \! i [\gamma^\mu, \gamma^\nu]/2$, and this term gives rise to the nucleon electric dipole moment \cite{Crewther:1979pi,Pospelov:1999mv,Chupp:2017rkp,Graham:2013gfa}
\beq
d_n = g_{aD} \, a \approx 2.4 \times 10^{-16} \frac{a}{f_a} e  \cdot {\rm cm}.
\eeq
In the presence of an oscillating axion background field, $a(t)$, this interaction generates an oscillating nucleon EDM. Thus, when dark matter consists of the axion field with $\rho_{\rm DM} \approx m_a^2 \, a_0^2/2\approx 0.4$ GeV/cm$^3$ \cite{Catena:2009mf} it could lead to an observable effect on nucleons.

\begin{figure}[tbp]
\centering
\includegraphics[width=.85\textwidth]{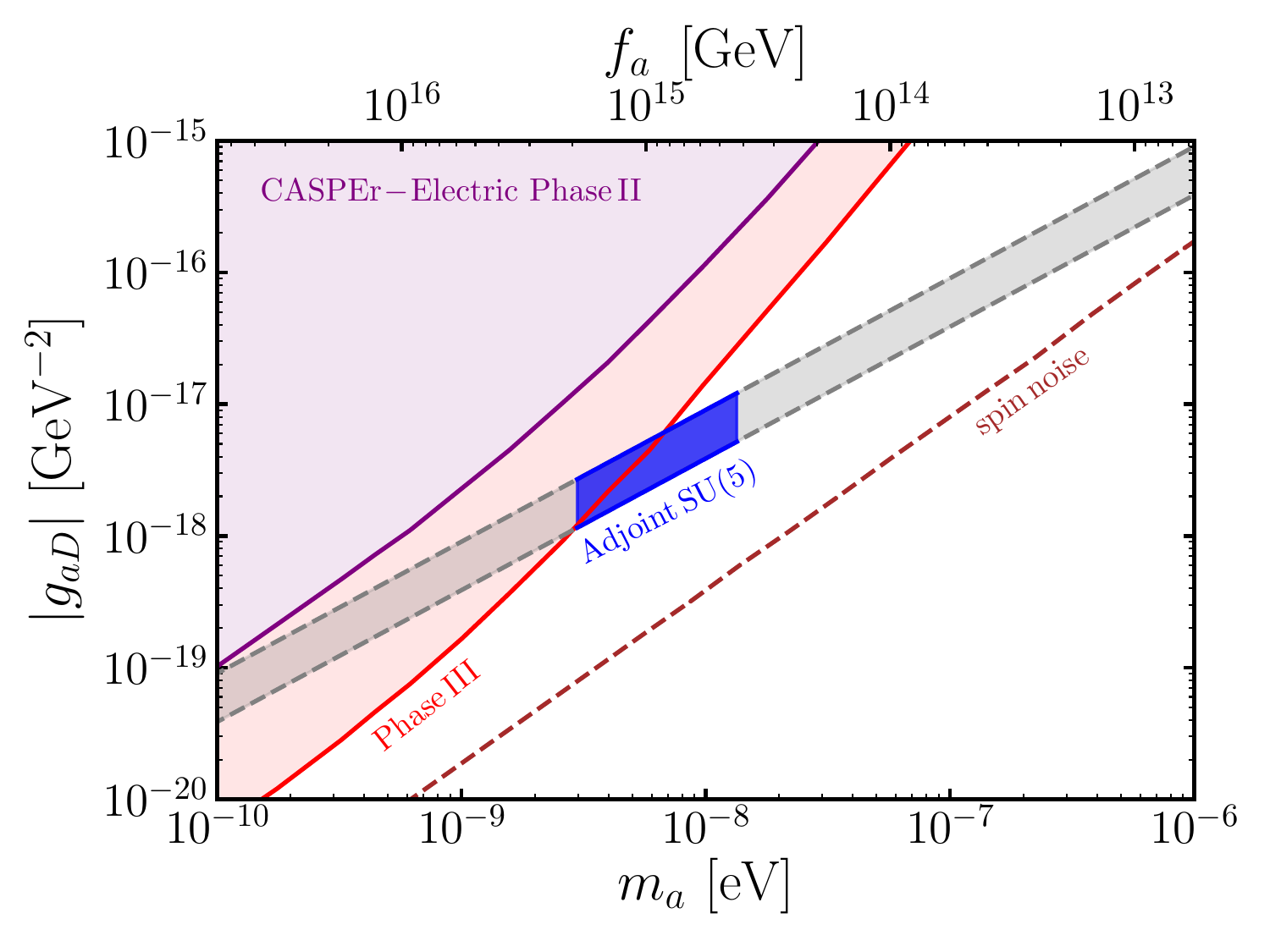}
\caption{\small{The axion coupling to the nucleon electric dipole moment versus the axion mass. The blue band is the prediction in Adjoint $\SU(5)$. The width of the band corresponds to the theoretical error, as given in \cite{Pospelov:1999mv,Chupp:2017rkp}. The region shaded in purple and red correspond to the projected sensitivities for Phase II and III of the CASPEr-Electric experiment \cite{JacksonKimball:2017elr}. The brown dashed line shows the ultimate limit from magnetization noise.}}
\label{fig:nEDM}
\end{figure}

The CASPEr-Electric experiment~\cite{Budker:2013hfa} has been proposed to measure this oscillating EDM by applying nuclear magnetic resonance techniques. This experiment is expected to probe axions with very low masses, $m_a < 10^{-9}$ eV, which is the region of interest for theories in which the PQ scale has a connection to the GUT or the string scale \cite{Svrcek:2006yi}.
In Fig.~\ref{fig:nEDM} we present the predictions for the $g_{aD}$ coupling in the axion mass window predicted by Adjoint $\SU(5)$. The region shaded in purple (red) corresponds to the projected sensitivity for Phase II (Phase III) for CASPEr-Electric \cite{JacksonKimball:2017elr}, the latter will be able to probe a part of the mass window predicted by Adjoint $\SU(5)$. The brown dashed line shows the ultimate limit from magnetization noise. As can be appreciated, these predictions for the axion mass can be probed in the near future.

In this article, we have showed that combining the signals at proton decay experiments, the collider signatures of the $\Phi_1$ field and the predictions for axion experiments, it is possible to test a simple grand unified theory which predicts a small window for the axion mass.
In this theory, we can also have an explanation for the baryon asymmetry in the Universe using the Baryogenesis via Leptogenesis mechanism. 
This theory predicts that $M_{\rho_0}= M_{\rho_3}/3$, then the lightest field relevant for leptogenesis is the field responsible for type I seesaw. The leptogenesis 
mechanism in the context of Adjoint $\SU(5)$ was investigated in Refs.~\cite{Blanchet:2008cj,Blanchet:2008ga}, but the emphasis was in the case where 
$\rho_3$ is the lightest field. We will investigate this issue in a future publication.
%\newpage
%%%%%%%%%%%%%%%%
\section{Summary}
\label{sec:Summary}
%%%%%%%%%%%%%%%

The QCD axion remains one of the most appealing candidates for the dark matter in the Universe. 
In this work, we discussed the implementation of the DFSZ and KSVZ mechanisms in renormalizable 
grand unified theories based on the gauge symmetry $\SU(5)$. In order to connect the GUT scale to the Peccei-Quinn scale we require the same scalar field to break both symmetries. We showed that in the simplest renormalizable models the ${\bf 5_H}$ and ${\bf 45_H}$ Higgses must couple in the same way to matter and have the same Peccei-Quinn charge. 
Therefore, the DFSZ mechanism cannot be implemented with the minimal Higgs sector even if we have two Higgs doublets and the singlet inside the ${\bf 24_H}$. 

We have shown that in the context of Adjoint $\SU(5)$ the same fermionic field, ${\bf 24}$, 
needed to generate neutrino masses can be used to implement the KSVZ mechanism. The colored and the charged fields in the ${\bf 24}$ generate the couplings of the axion to the gluons and the photons respectively, while neutrino masses are generated through the seesaw mechanism. We found that the theory studied in this article can be considered as the realistic renormalizable theory based on $\SU(5)$ with the minimal number of representations where the Peccei-Quinn mechanism can be realized. 

We investigated the unification of gauge interactions taking into account collider and proton decay constraints, 
and we found that the allowed GUT scale is in a small window which allows us to predict the axion mass in the range
$$m_a \simeq (3 - 13) \times  10^{-9} \  \rm{eV}.$$ 
We focused on the scenario where the PQ symmetry is broken before inflation and discussed the predictions for the different axion experiments showing that this theory could be fully tested in the near future by the ABRACADABRA and the CASPEr-Electric experiments.
Additionally, we have shown that this theory could be probed at proton decay experiments such as the current Super-Kamiokande and the future Hyper-Kamiokande.

The prediction for the axion mass in this work has some overlap with the one given in Ref.~\cite{DiLuzio:2018gqe}. In the latter work the authors rely on Planck suppressed operators that could break the PQ symmetry 
and used two mechanisms to generate the axion mass . In our case, we have considered a simple renormalizable theory that does not rely on Planck suppressed operators and the axion mass is generated only by the KSVZ mechanism. 
Our theory is more predictive due to renormalizability and it has less free parameters.

The theoretical framework presented in this article, based on the $\SU(5) \times \U(1)_{\rm PQ}$ symmetry, can be considered as an appealing theory for physics beyond the Standard Model, where it is possible to understand the unification of the Standard Model forces, the origin of neutrino masses, dark matter, the strong CP problem and the baryon asymmetry through leptogenesis. Furthermore, the lifetime of the proton can be predicted and exotic signatures at colliders are expected.

\section*{Acknowledgments}
P. F. P. thanks M. B. Wise for a discussion and the theory group at Caltech for hospitality.
A. D. P. would like to thank C. Tamarit for helpful discussions. The work of C. M. has been supported in part by Grants
No. FPA2014-53631-C2-1-P, No. FPA2017-84445-P, and
No. SEV-2014- 0398 (AEI/ERDF, EU), and by a La Caixa-Severo Ochoa scholarship.

\appendix

%%%%%%%%%%%%%%%%%%%%%%
\section{Adjoint $\SU(5)$ Representations}
\label{sec:AppRep}
%%%%%%%%%%%%%%%%%%%%%%
In the Adjoint $\SU(5)$~\cite{Perez:2007rm} theory we have the following representations: 

\begin{equation*}
{\bf{\bar{5}}} = \begin{pmatrix} d^c_1 \\  d^c_2 \\ d^c_3 \\ e \\ - \nu \end{pmatrix},  
\\  
\quad  \quad 
\\ 
{\bf{10}} = \frac{1}{\sqrt{2}} \begin{pmatrix} 0 & u^c_3 & -u^c_2 & - u^1 & -d^1   \\ 
							 -u^c_3 &  0 & u_1^c & - u^2 & - d^2 \\ 
							 u^c_2 & -u^c_1 & 0 & - u^3 & - d^3 \\
							 u^1 & u^2 & u^3 & 0 & e^c \\
							 d^1 & d^2 & d^3 & - e^c & 0 
							 \end{pmatrix}, 
\\  
\quad  \quad 
\\
{\bf{5}_H} = \begin{pmatrix} T_1 \\  T_2 \\  T_3 \\ H^+ \\ H^0  \end{pmatrix}, 
\end{equation*}

\begin{equation*}
{\bf{24_H}} = \begin{pmatrix} \Sigma_8 - \frac{2}{\sqrt{30}}\Sigma_0 & \Sigma_{(\bar{3},2)} 
\\ \Sigma_{({3},2)} & \Sigma_3 + \frac{3}{\sqrt{30}}\Sigma_0 \end{pmatrix}, 
\\  
\quad  \quad 
\\
{\bf{24}} = \begin{pmatrix} \rho_8 -\frac{1}{\sqrt{15}}\rho_0 & \rho_{({\bar{3}},2)} \\ \rho_{({3},2)} & \rho_3 + \frac{3}{2\sqrt{15}} \rho_0 \end{pmatrix}, 
\end{equation*}
\newline
\noindent the components of the ${\bf{24_H}}$ have the following quantum numbers,
$$\Sigma_8 \sim (8,1,0), \,\, \Sigma_3 \sim (1,3,0), \,\, \Sigma_0 \sim (1,1,0),  \,\, \Sigma_{({3},2)} \sim (3,2,-5/6) \   {\rm{and}} \  \Sigma_{(\bar{3},2)} \sim (\bar{3},2,5/6).$$
In a similar manner, the components of the ${\bf{24}}$ are given by
$$\rho_8 \sim (8,1,0), \,\, \rho_3 \sim (1,3,0), \,\,  \rho_0 \sim (1,1,0), \,\,  \rho_{({3},2)} \sim (3,2,-5/6) \   {\rm{and}} \  \rho_{(\bar{3},2)} \sim (\bar{3},2,5/6), $$
while the gauge bosons are described by
\begin{equation*}
{\bf{A}_\mu} = \begin{pmatrix} G_\mu - \frac{1}{\sqrt{15}} B_\mu & V_\mu^\dagger 
\\ 
V_\mu & W_\mu + \frac{3}{2\sqrt{15}} B_\mu \end{pmatrix}, 
\end{equation*}
with quantum numbers
$$G_\mu \sim (8,1,0), \  B_\mu \sim (1,1,0), \ V_\mu \sim (3,2,-5/6) \  {\rm{and}} \ W_\mu \sim (1,3,0).$$
The scalar 45 dimensional representation has components
\begin{eqnarray*}
%45 Higgses
 \displaystyle {\bf{45_H}} &\sim&\underbrace{(8,2,1/2)}_{\Phi_1}\oplus \underbrace{(\bar{6},1,-1/3)}_{\Phi_2}\oplus \underbrace{(3,3,-1/3)}_{\Phi_3}
 \oplus \underbrace{(\bar{3},2,-7/6)}_{\Phi_4}\oplus \underbrace{(3,1,-1/3)}_{\Phi_5} \oplus \underbrace{(\bar{3},1,4/3)}_{\Phi_6}
 \oplus \underbrace{(1,2,1/2)}_{H_2}. \\ \label{45H}
\end{eqnarray*}
\newpage

%%%%%%%%%%%%%%%%%%%%%%
\section{Proton Decay}
\label{sec:AppProton}
%%%%%%%%%%%%%%%%%%%%%%

The proton decay widths for the most relevant channels are given by 
\begin{eqnarray}
\Gamma (p \to \pi^0 e^+ )&=& \frac{m_p}{8 \pi} A^2 k_1^4 \left(  \left| c(e^c,d) \matrixel{\pi^0}{(ud)_L u_R}{p}\right |^2  + \left| c(e,d^c) \matrixel{\pi^0}
{(ud)_R u_L}{p}\right|^2 \right), \hspace{0.8cm} \\
\displaystyle \Gamma (p \to K^+ \bar{\nu}) &=& \frac{m_p}{8 \pi} \left( 1- \frac{m_{K^+}^2}{m_p^2}\right)^2 A^2 k_1^4 \sum_i  \left|  
c(\nu_i,d,s^c)  \matrixel{K^+}{(us)_R d_L}{p}  \right. \nonumber \\
& & + \left. c(\nu_i,s,d^c)  \matrixel{K^+}{(ud)_R s_L}{p} \right|^2,\\[1ex]
\Gamma (p \to \pi^+ \bar{\nu}) &=& \frac{m_p}{8 \pi} A^2 k_1^4\left|V_{UD}^{11} \matrixel{\pi^+}{(du)_R d_L}{p}\right |^2,\\  \nonumber  \\
\textrm{with} \ A&=&A_{\rm QCD} A_{SR}=\left( \frac{\alpha_3 (m_b)}{\alpha_3 (M_Z)} \right)^{6/23} \left( \frac{\alpha_3 (Q)}{\alpha_3 (m_b)} \right)^{6/25} \left( \frac{\alpha_3 (M_Z)}{\alpha_3 (M_{\rm GUT})} \right)^{2/7}.
\end{eqnarray}
In the above equation $k_1=g_{\rm GUT}/\sqrt{2} M_{\rm GUT}$ and $A$ defines the running of the operators. $A_{\rm QCD}\approx 1.2$ corresponds to the running from the $M_Z$ to the $Q\approx 2.3$ GeV scale, while $A_{SR} \approx 1.5$ defines the running from the GUT scale to the electroweak scale. 
The c-coefficients~\cite{FileviezPerez:2004hn} are given by
\begin{eqnarray}
&& c(e^c_\alpha, d_\beta) = V_1^{11} V_2^{\alpha \beta} + (V_1 V_{UD})^{1 \beta} (V_2 V_{UD}^\dagger)^{\alpha 1},\\
&& c(e_\alpha, d^c_\beta) = V_1^{11} V_3^{\beta \alpha},\\
&& c(\nu_l,d_{\alpha},d_{\beta}^c) = (V_1V_{UD})^{1\alpha}(V_3 V_{EN})^{\beta l},
\end{eqnarray}
where the $V$'s are mixing matrices defined as 
\begin{eqnarray}
V_1=U_C^{\dagger}U, \,\, V_2=E_C^{\dagger}D, \,\, V_3=D_C^{\dagger}E, \,\,
V_{UD}=U^{\dagger}D \  \textrm{and} \  V_{EN}=E^{\dagger}N. 
\end{eqnarray}
The matrices $U$, $E$, $D$ and $N$ define the Yukawa couplings diagonalization so that 
\begin{eqnarray}
U_C^T Y_u U=Y_u^{\text{diag}}, \,\,
D_C^T Y_d D=Y_d^{\text{diag}}, \,\,
E_C^T Y_e E=Y_e^{\text{diag}},\ \textrm{and} \
N^T Y_\nu N=Y_\nu^{\text{diag}}.
\end{eqnarray}
For the predictions of the proton decay width we have taken the values for the hadronic matrix elements given in Ref.~\cite{Aoki:2017puj}:
\begin{align}
&\matrixel{\pi^0}{(ud)_Ru_L}{p}=-0.131 \text{ GeV}^2, \quad & \matrixel{\pi^+}{(du)_R d_L}{p}=-0.186 \text{ GeV}^2, \nonumber\\
&\matrixel{K^+}{(us)_R d_L}{p}=-0.049 \text{ GeV}^2, \quad  &\matrixel{K^+}{(ud)_R s_L}{p}=-0.134 \text{ GeV}^2.  \nonumber
\end{align}

%
%%%%%%%%%%%%%%%%%%%%%%
\section{BBN bounds on the $\rho_8$ lifetime}
\label{sec:AppBBN}
\begin{figure}[tbp]
\centering
\includegraphics[width=.6\textwidth]{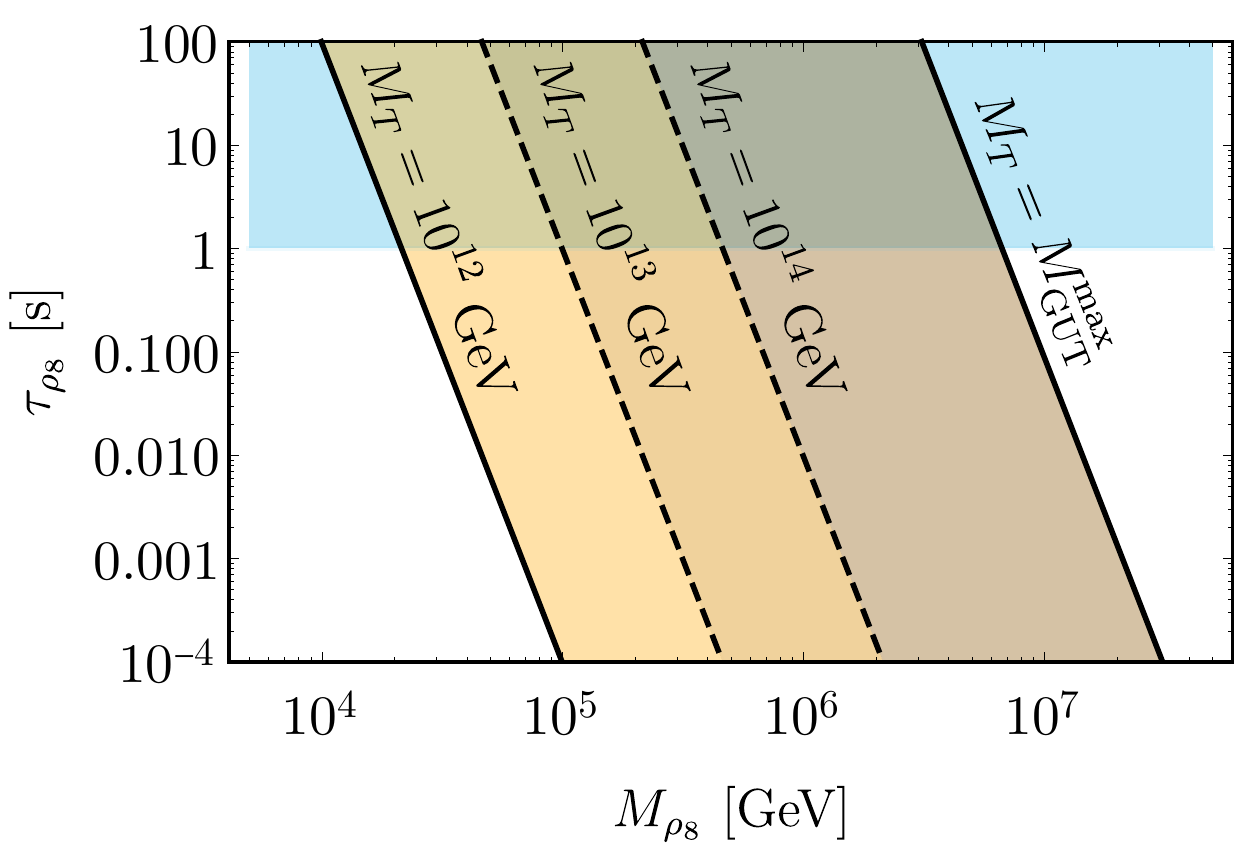}
\caption{Predictions of the $\rho_8$ lifetime for different $M_T$ masses, we have $M_T>10^{12}$ GeV from bounds on the proton lifetime. The blue shaded regions shows the predictions for $ \tau_{\rho_8}\!<\!1 \ {\rm{sec}}$ and excluded by a naive bound coming from Big Bang Nucleosynthesis.}
\label{rho8}
\end{figure}

%%%%%%%%%%%%%%%%%%%%%%
%
We have discussed above the unification constraints and mentioned that the allowed value of the gauge coupling  at the GUT scale is sensitive to 
the mass of the fields in the ${\bf{24}}$ representation.  The fields $\rho_0$ and $\rho_3$ have two body decays and can decay 
before Big Bang Nucleosynthesis (BBN), while the colored fields $\rho_8$, $\rho_{(3,2)}$ and $\rho_{(\bar{3},2)}$ have only three body decays.
Here we discuss only the bounds on the mass of $\rho_8$ by imposing that the decay lifetime must be smaller than one second to avoid issues with BBN~\cite{Fields:2014uja}. 
Since the masses of all fields in ${\bf{24}}$ are related it is enough to understand the limit on the mass of $\rho_8$. Neglecting all 
SM fermion masses, the decay width for $\rho_8$ decaying into three quarks is given by
\begin{equation}
\Gamma(\rho_8) \simeq \frac{Y_t^2 h_1^2 }{48 (2\pi)^3}\frac{M_{\rho_8}^5}{M_T^4},
\end{equation}
where $M_T$ corresponds to the mass of the scalar triplet inside the ${\bf 5_H}$. Here we used the main decay channel defined by the coupling of $T$ to the third generation quarks.
Naively, the $\rho_3$ contribution to neutrino masses can be written as
\begin{equation}
\label{eq:nuseesaw}
M_\nu \simeq  \frac{h_1^2v_0^2}{M_{\rho_3}},
\end{equation}
and using this result we find that
\begin{equation}
\Gamma(\rho_8) \simeq \frac{Y_t^2 }{32 (2\pi)^3}\frac{M_\nu}{v_0^2}\frac{M_{\rho_8}^6}{M_T^4}.
\end{equation}
In Fig.~\ref{rho8} we present the predictions for the $\rho_8$ lifetime using different $M_T$ masses ($M_T > 10^{12}$ GeV from bounds on the proton lifetime) and $M_\nu \sim 0.1$ eV. 
The region shaded in blue shows the region that leads to $ \tau_{\rho_8}  > 1 \ {\rm{sec}}$ and it is excluded by a naive bound coming from Big Bang Nucleosynthesis. Therefore, the mass of $\rho_3$ should lie in the window
\beq
10^{4.51} \,\, {\rm GeV} < M_{\rho_3} < 10^{15} \,\, {\rm GeV},
\eeq
where the latter bound comes from the perturbativity of the Yukawa coupling in Eq.~\eqref{eq:nuseesaw}.

%%%%%%%%%%%%%%%%%%%%%%
\newpage
\section{Beta Functions}
\label{sec:AppBeta}
%%%%%%%%%%%%%%%%%%%%%%
The extra contributions to the running of the gauge couplings in Adjoint $\SU(5)$ are listed in Table~\ref{tab:beta}.
\begin{table}[ht]
\centering
\begin{tabular}{lcccc}
 \text{Fields} & $b_1$ & $b_2$ & $b_3$
\\
\hline
\hline
$T$ &  1/15 & 0 & 1/6
\\
\hline
$\Phi_{1}$ &  4/5 & 4/3 & 2
\\
\hline
$\Phi_{2}$ &  2/15 & 0 & 5/6
\\
\hline
$\Phi_{3}$ &  1/5 & 2 & 1/2
\\
\hline
$\Phi_{4}$ & 49/30 &  1/2 & 1/3
\\
\hline
$\Phi_{5}$ & 1/15 & 0 & 1/6
\\
\hline
$\Phi_{6}$ &  16/15 & 0 & 1/6
\\
\hline
$H_2$ & 1/10 & 1/6 & 0
\\
\hline
$\rho_8$ &  0 & 0 & 2
\\
\hline
$\rho_3$ & 0 & 4/3 & 0
\\
\hline
$\rho_{(3,2)}$   & 5/3 & 1 & 2/3 \\
\hline
$\rho_{(\bar{3},2)}$  & 5/3 & 1 & 2/3 \\
\hline
\end{tabular}
\caption{\label{tab:beta} Contributions from each field to the renormalization group equations for the gauge couplings.}
\end{table}

The $B_{ij}$ parameters that enter in Eqs.~\eqref{unificationCONS1}-\eqref{unificationCONS2} correspond to $B_{ij}=B_i - B_j$ with
\beq
B_i = b_i^{\rm SM} + \sum_I b_{iI}  \, \frac{ \ln \left( M_{\rm GUT} / M_I \right) }{ \ln \left( M_{\rm GUT} / M_Z \right)  },
\eeq
where $M_I$ is the mass of the new particle living between the electroweak and the GUT scales.
%%%%%%%%%%%%%%%%%%

\bibliographystyle{JHEP}
\bibliography{axions}{}
\end{document}